\documentstyle[12pt,WSPRL]{article}

\begin{document}

\centerline{\normalsize\bf BILEPTONS - STATUS AND PROSPECTS } 

\baselineskip=16pt
\vspace*{0.3cm}

\centerline{\footnotesize PAUL H. FRAMPTON}
\baselineskip=13pt
\centerline{\footnotesize\it Department of Physics and Astronomy,}
\baselineskip=12pt
\centerline{\footnotesize\it University of North Carolina, Chapel Hill, NC 27599-3255}
\centerline{\footnotesize E-mail: frampton@physics.unc.edu}
\vspace*{0.9cm}

\abstracts{ 
Theoretical backgound for bileptonic gauge bosons is reviewed, both
the SU(15) GUT model and the 3-3-1 model. Then the mass limits for
bileptons are discussed coming from $e^+e^-$ scattering, polarized
muon decay and muonium-antimuonium conversion. 
A consequence of precision electroweak data on the masses is mentioned.
Discovery in $e^-e^-$ at a linear collider is emphasised.}
\bigskip
\bigskip
\noindent

\bigskip
{\bf Theoretical Background.}
\bigskip

The simplest grand unified theory (GUT) is SU(5) where
each family is $(10 + \bar{5})_L$. The gauge bosons mediate proton
decay with (in the minimal form) too fast a rate. The
gauge bosons have no definite $B$ or $L$, only
$(B - L)$ which is conserved.

SU(15) arose from the idea of having unification where the gauge bosons
have well-defined $B$ (and $L$) and so do {\it not}
mediate proton decay. Each family is assigned to a {\bf 15} of
SU(15); each of the {\bf 224} gauge bosons then have
a definite $B$ and $L$. Hence proton decay is absent in the gauge
sector.

The first stage of symmetry breaking takes SU(15) to
$SU(12)_q \times SU(3)_l$ at a GUT scale $M_G$.
$SU(3)_l$ acts on the lepton triplet
$(e^+, \nu_e, e^-)$ and contains
the SU(2) doublet of gauge bosons
$(Y^{--}, Y^-)$ and the antiparticles $(Y^{++}, Y^+)$.
These bileptons have a mass at or just above the weak scale,
say in the region $300$GeV to $600$GeV (this can be made
more rigorous). In particular, the idea that a narrow
resonance might appear in
$e^-e^- \rightarrow \mu^-\mu^-$ was suggested in \cite{FL}.
The width is predicted as a few per cent of the mass.

The anomaly cancellation of SU(15) is inelegant: by mirror
fermions. This is an aesthetic consideration - not
a phenomenological one. Nevertheless, it is interesting to
know that there is a chiral model which incorporates the
bileptons.

To introduce the 331 Model\cite{PHF} the following are motivating factors:

i) Consistency of a gauge theory (unitarity, renormalizability) requires
anomaly cancellation. This requirement almost alone is able to fix all
electric charges and other quantum numbers within one family
of the standard model. This accounts for charge quantization, 
{\it e.g.} the neutrality of
the hydrogen atom, without the need for a GUT.

ii)This does not explain why $N_f > 1$ for the number of families but is
sufficiently impressive to suggest that $N_f = 3$ may be explicable by
anomaly cancellation in an extension of the standard model. 
This requires that each
extended family have non-vanishing anomaly and that the three families 
are not all treated
similarly.

iii) A striking feature of the mass spectrum in the SM is the top mass suggesting that
the 3rd. family be treated differently and that the anomaly cancellation be proportional
to: +1 +1 -2 = 0.  

iv)There is a " -2 " lurking in the SM in the ratio of the quark electric charges!

v)The electroweak gauge group extension from $SU(2)$ to $SU(3)$ will add five
gauge bosons. The adjoint of $SU(3)$ breaks into $8 = 3 + (2 + 2) + 1$ under $SU(2)$.
The $1$ is a $Z^{'}$ and the two doublets are readily identifiable from the leptonic triplet
or antitriplet $(e^{-}, \nu_e , e^{+})$ as {\it bilepton} gauge bosons
$( Y^{--} , Y^{-})$ with $L = 2$ and $( Y^{++} , Y^{+})$ with $L = -2$.
Such bileptons appeared first in stable-proton GUTs but there the fermions
were non-chiral and one needed to invoke mirror fermions; this is
precisely what is avoided in the 331 Model.  But it is true that the $SU(3)$
of the 331 Model has the same couplings to the {\it leptons} as that of the leptonic
$SU(3)_l$
subgroup of $SU(15)$ which breaks to $SU(12)_q \times SU(3)_l$  .

\bigskip

Now I am ready to introduce the 331 Model in its technical details: the gauge group
of the standard model is extended to $SU(3) \times SU(3) \times U(1) $ where the electroweak
$SU(3)$ contains the standard $SU(2)$ and the weak hypercharge is a mixture of $ \lambda_8 $
with the $U(1)$. The leptons are in the antitriplet $( e^-, \nu_e, e^+ )_L$ and similarly
for the $\mu$ and $\tau$.

These antitriplets have $X = 0$ where $X$ is the new $U(1)$ charge. This can be checked by
noting that the $X$ value is the electric charge of the central member of the triplet or
antitriplet.

For the first family of quarks I use the triplet $( u, d, D )_L$ with $X = -1/3$
and the right-handed counterparts in singlets. Similarly, the second family of
quarks is treated. For the third family of quarks, on the other hand, I use the
{\it antitriplet} $( T, t, b)_L$ with $X = +2/3$. The new exotic quarks D, S, and T have
charges -4/3, -4/3 and +5/3 respectively.

It is instructive to see how this combination successfully cancels all chiral
anomalies:

The purely color anomaly $(3_L)^3$ cancels because QCD is vector-like.

The anomaly $(3_L)^3$ is non-trivial. Taking, for the moment, arbitrary
numbers $N_c$ of colors and $N_l$ of light neutrinos I find this anomaly cancels only
if $N_c = N_l = 3$.

The remaining anomalies $(3_c)^2X$, $(3_L)^2X$, $X^3$ and $X(T_{\mu\nu}^2$
also cancel.

Each family separately has non-zero anomaly for $X^3$, $(3_L)^2X$ and $(3_L)^3$;
in each case, the anomalies cancel proportionally to $+1 +1 -2$ between the 
families.

\medskip

To break the symmetry I need several Higgs multiplets. A triplet $\Phi$ with $X = +1$
and VEV $<\Phi>$ = $(0,0,U)$ breaks 331 to the standard 321 group, and gives masses to
D, S, and T as well as to the gauge bosons Y and Z'.  The scale U sets the range of the 
new physics and I shall discuss more about its possible value.

The electroweak breaking requires two further triplets $\phi$ and $\phi'$ with 
$X = 0$ and $X = -1$ respectively. Their VEVs give masses to d, s, t and to
u, c, b respectively. The first VEV also gives a contribution of an
antisymmetric-in-family type to the charged leptons. To complete a satisfactory
lepton mass matrix necessitates adding a sextet with $X = 0$. 

\medskip

What can the scale $U$ be? It turns out that there is not only the lower
bound expected from the constraint of the precision electroweak data,
but also an upper bound coming from a group theoretical constraint within
the theory itself.

The lower bound on $U$ from $Z-Z'$ mixing can be derived from the
diagonalization of the mass matrix and leads to $M(Z') \geq 300GeV$.
The limit from FCNC (the Glashow - Weinberg rule is violated) gives a similar 
bound; here the suppression is helped by ubiquitous $(1 - 4sin^2\theta)$
factors.

In these considerations, particularly with regard to FCNC, the special role
played by the third family is crucial; if either of the first two families
is the one treated asymmetrically the FCNC disagree with experiment.

\medskip

The upper bound on $U$ arises because the embedding of the standard $321$
group in $331$ requires that $sin^2\theta \leq 1/4$. When $sin^2\theta = 1/4$,
the $SU(2) \times U(1)$ group embeds entirely in $SU(3)$, and the coupling of the $X$ charge in principle diverges. Because the phenomenological value is close to 1/4 -
actually $sin^2\theta (M_Z) = 0.233$ - the scale $U$ must be less than about
$3TeV$ after scaling $sin^2\theta(\mu)$ by the renormalization group. Putting some
reasonable upper bound on the $X$ coupling leads to an upper bound on the bilepton
mass, for this 331 Model, of about $800GeV$ [ Here I have allowed one further Higgs multiplet
- an octet].  
\bigskip

{\bf Allowed Masses}

\bigskip

\medskip

In order to set limits on the minimum mass allowed by the experimental data,
a number of processes were considered initially\cite{FN}.
AT the time (1992), as now, the most accurate
high-energy data on the planet came from LEP. Analysis of $e^+e^-$ 
scattering at LEP was used to give
a lower limt of about 120GeV on the singly-charged bilepton, from
the fact that its exchange in the u-channel effects
angular distributions observed. This limit which did seem surprisingly weak was
the best found at the time.

The most useful experiment for limiting the singly-charged
bilepton mass from below is polarized muon decay. 
This is an example where a {\it very} low energy experiment
with sufficient precision can compete successfully
with the highest energy experiments
in limiting the possible properties of the highest
accessible mass particles.
With the coupling parametrized as $V - \xi A$
where $\xi$ is a Michel parameter, 
the present limit\cite{muon}
on $\xi$ is $1 \geq \xi \geq 0.997$
coming from about $10^8$ examples of the decay. 
This leads to a lower bound
$M(Y^{\pm}) \geq 230$GeV for the singly-charged bilepton,
from this direct measurement. 
Since 
\begin{equation}
(1 - \xi ) \sim (M_W/M_Y)^4
\end{equation}
I deduce that if $(1 - \xi )$ could be measured to an accuracy of $10^{-4}$
the limit would become $M_Y \geq 10 M_W$ and if to an accuracy $10^{-8}$
it would be $M_Y \geq 100M_W$. The first of these is within the realm
of feasability and certainly seems an important experiment to pursue.
The group at the Paul Scherrer Institute near Zurich (Gerber, Fetscher)
is one that is planning this experiment.

A new limit since the previous Santa Cruz $e^-e^-$ workshop
in 1995 comes\cite{abela} from muonium-antimuonium conversion
which provides a lower limit $M(Y^{\pm\pm}) > 360$GeV
on the doubly-charged bilepton.

\bigskip
By studying the $S$ and $T$ parameters which meaure\cite{PT}
the compatibility of theory with high precion
experiment, it can be shown\cite{FH}
that given the lower mass bound on the doubly-charged
bilepton, the singly-charged bilepton
must have a mass $M(Y^{\pm}) > 320$GeV, a tighter bound
than the direct measurement. 

The upper mass limit from the theory with minimal Higgs in the 331-model,
is 
$M(Y^{\pm}, Y^{\pm\pm}) < 600$GeV. 
As mentioned above,
if one extends the Higgs structure by adding an octet of $SU(3)_L$
this limit relaxes to
$M(Y^{\pm}, Y^{\pm\pm}) < 800$GeV.
So this is somewhat model-dependent. 

A careful analysis of all the mass limits on bileptons has
been carried out in the last year by Cuypers\cite{cuypers}.

The crucial question is how to produce the bilepton
in experiment.
The bilepton can be produced in a hadron collider such as a $pp$ or $p \overline{p}$
machine, or in a lepton collider such as $e^+e^-$ or $e^-e^-$.

For the hadron collider the $Y$ may be either pair produced or produced in association
with an exotic quark [the latter carries $L = \pm 2$]. It turns out that the associated
production is about one order of magnitude larger. These cross-sections are
calculated in the literature - for a pp collider of the type envisioned there would
be at least $10^4$ striking events per year.

\medskip

Surely the most dramatic way to spot a bilepton, however, would be to run a linear collider in
the $e^-e^-$ mode and find a direct-channel resonance. A narrow spike at between $300GeV$
and $1000GeV$ would have a width at most a few percent of its mass and its decay to $\mu^-\mu^-$ has no standard model background.

The discovery of such a bileptonic gauge boson will strongly suggest
that the electroweak isospin $SU(2)$ is to be regarded as
a subgroup of $SU(3)_L$. The situation can be regarded
as roughly analogous to the extension of nuclear
isospin to the flavor SU(3); the bileptons play the role
in that analogy of the K meson doublets filling out the adjoint
octet of SU(3). The shortcoming of that analogy (unfortunately!)
is that there the strange particles had been already
discovered {\it before} the theory was extended.

\bigskip

{\bf Acknowledgement}

\bigskip

This work was supported in part by the U.S. Department of Energy under Grant DE-FG05-85ER-40219.\\


\bigskip
\bigskip
\newpage


\bigskip


{\bf References}

\bigskip

\begin{thebibliography}{999}
\bibitem{FL}
P. H. Frampton and B. H. Lee, Phys. Rev. Lett. {\bf 64}, 619 (1990). 

\bibitem{PHF} 
P. H. Frampton, Phys. Rev. Lett. {\bf 69}, 2889 (1992);

\bibitem{FN}
P.H. Frampton and D. Ng, Phys. Rev. {\bf D45,} 4240 (1992).

\bibitem{muon}
A. Jodidio {\it et al.} Phys. Rev. {\bf D34,} 1967 (1986);
Err. {\bf D37,} 237 (1988).

\bibitem{abela}
R. Abela {\it et al.} Phys. Rev. Lett. {\bf 77,} 1950 (1996).

\bibitem{PT}
M.E. Peskin and T. Takeuchi, Phys. Rev. Lett. {\bf 65,} 2963 (1990).

\bibitem{FH}
P.H. Frampton and M. Harada, UNC-Chapel Hill Reports IFP-746-UNC
and IFP-748-UNC (1997).      

\bibitem{cuypers}
F. Cuypers and S. Davidson, PSI-PR-96-21 (1996) {\it hep-ph/9609487}\\
F. Cuypers and M. Raidal, Nucl. Phys. {\bf B501,} 3 (1997).

\end{thebibliography}
\end{document}